\documentclass[12pt]{article} 

\usepackage{mathtools, amsfonts, amsthm, latexsym, amssymb,dsfont}
\usepackage[T1]{fontenc}
\usepackage[tracking=true]{microtype}
\usepackage{lmodern}

\usepackage[margin=1in]{geometry}

\usepackage{graphicx}
\usepackage{subcaption}
\usepackage{booktabs}
\usepackage{longtable}
\usepackage{cite}

\usepackage{color}
\definecolor{darkblue}{cmyk}{0.9,0.9,0,0}
\definecolor{wine-stain}{rgb}{0.5,0,0}
\usepackage[colorlinks, allcolors=darkblue, linktoc=all]{hyperref} 
\renewcommand{\d}{\partial}
\renewcommand{\O}{\mathcal{O}}
\renewcommand{\L}{\mathcal L}

\newcommand{\vev}[1]{{\langle #1 \rangle}}
\newcommand{\del}{\nabla}
\newcommand{\CPN}{\mathbb{C}P^{N-1}}
\DeclareMathOperator{\tr}{tr}

\begin{document}
\thispagestyle{empty}
\vspace*{1in} 

\begin{center}
{\Large \textbf{The large charge expansion at large $N$}}
\vspace{.6in}

\textbf{Anton de la Fuente} 
\vspace{.2in} 

\textit{Theoretical Particle Physics Laboratory, Institute of Physics, EPFL \\  Lausanne, Switzerland} 
\end{center}
\vspace{.2in}

\begin{abstract}
The scaling dimensions of charged operators in conformal field theory have recently been predicted to exhibit universal behavior in the large charge limit. We verify this behavior in the 2+1 dimensional $\CPN$ model.
Specifically, we numerically compute the scaling dimensions of the lowest dimension monopole operators with charges $Q= 1,2,\cdots, 100$ to subleading order in large~$N$. The coefficients of the large~$Q$ expansion are extracted through a fit, and the predicted universal $\O(Q^0)$ contribution is verified to the subpercent level. 
\end{abstract}

\newpage
\section{Introduction}
Most observables in a generic quantum field theory cannot be computed due to the lack of a small expansion parameter. Some of the biggest breakthroughs in theoretical physics followed from the discovery of small parameters in theories that naively did not have any small parameters\cite{wilsonFisher, wilsonNandEpsilon,largeNoriginal,banksZaks,maldacenaOriginal}. This allowed for the computation of generic observables in a restricted class of theories. There is also the converse approach---effective field theory (EFT)---that instead focuses on the calculation of a restricted class of observables in generic theories. Different EFTs exist for different classes of observables (e.g.\ \cite{chiral, hqet, scet, nucleon}).

Recently, a new class of observables was found to be computable in a generic conformal field theory~(CFT) via EFT techniques \cite{hellerman1, alvarezGaume, riccardo, loukas, hellerman2, hellerman3, monteCarlo, matrixModels, hellerman4, zhiboedov, vortex, loukas2, hellerman5, loukas3, hellerman6}. When two of the operators in an $n$-point function have a large global charge, the correlator becomes computable in a large charge expansion. The 2-point function then leads to nontrivial predictions about operator scaling dimensions. Specializing to the case of a $U(1)$ global charge in 3~spacetime dimensions, the dimension~$\Delta$ of the lowest dimension operator with charge $Q$ is predicted to take the universal form
\begin{equation} \label{largeQ}
	\Delta = \alpha Q^{3/2} + \beta \sqrt Q + \gamma + \O\left(\frac1{\sqrt Q} \right),
\end{equation}
where $\alpha$ and $\beta$ are non-universal constants depending on the specific CFT, while
\begin{equation} \label{prediction}
	\gamma = -0.0937256\dots
\end{equation}
is a universal constant independent of the CFT \cite{hellerman1, sasha}.

The purpose of this paper is to verify this universal constant \eqref{prediction} in one of the CFTs that also has a small expansion parameter: the $\CPN$ model at large $N$  \cite{wittenCPN, colemanBook, zinnJustin, largeN1, largeN3, pufu}.  We use the methods of \cite{pufu} to numerically compute $\Delta$ to subleading order in $N$---but nonperturbative in~$Q$---for $Q = 1,2,\cdots, 100$. We then fit the coefficients of the large charge expansion~\eqref{largeQ} to our computed values of $\Delta$ and verify \eqref{prediction} to the subpercent level. 


\section{Monopole operators}
The charged operators that we will study are the monopole operators of a 2+1-dimensional~$U(1)$ gauge theory. To review, in 2+1 spacetime dimensions, any $U(1)$ gauge theory has a $U(1)$ global symmetry generated by the current
\begin{equation} \label{current}
	J_\mu = \tfrac{1}{4\pi} \epsilon_{\mu \nu \lambda} F^{\nu \lambda},
\end{equation}
where $F$ is the field strength of the gauge field. Local operators charged under this current are called monopole operators.  As a simple example, in free Maxwell theory, the $U(1)$ gauge field can be rewritten in terms of a free compact scalar $\pi \equiv \pi + 2\pi$ by writing $J_\mu = f\d_\mu \pi$ where~$J_\mu$ is as just defined \eqref{current} and $f$ is an arbitrary scale. The operator $e^{i Q \pi}$ is then a monopole operator of charge $Q$.

More generally, monopole insertions can be defined by restricting the path integral to gauge field configurations that satisfy
\begin{equation} \label{flux}
	\int_{S^2} F = 2\pi Q
\end{equation}
whenever the $S^2$ encloses a monopole of charge $Q$. This definition can be hard to use in practice since one also needs to supply boundary conditions for all the other fields. In fact, different boundary conditions are what distinguish different operators of charge $Q$.

We will instead define monopole operators by the state-operator correspondence~\cite{largeN2}. In particular, the lowest dimension monopole operator of charge $Q$ corresponds to the ground state of the CFT on the cylinder $S^2 \times \mathbb R$ in the sector with fixed magnetic flux \eqref{flux} (where the~$S^2$ in \eqref{flux} now refers to the $S^2$ of the cylinder $S^2\times \mathbb R$ instead of to an arbitrary $S^2$ embedded in $\mathbb R^3$). The energy of this state is the scaling dimension $\Delta$ of the operator, and $\Delta$ can be written in terms of the zero temperature limit of the partition function $Z_Q$ as
\begin{equation} \label{stateOp}
	\Delta = - \lim_{\beta \to \infty} \frac{1}{\beta} \ln Z_Q.
\end{equation}
The partition function $Z_Q$ only sums over states of charge $Q$. In other words, all gauge field configurations in the Euclidean path integral that computes $Z_Q$ are restricted to satisfy \eqref{flux}. In \eqref{stateOp} and for the rest of the paper, we set the radius $R$ of the $S^2$ to 1.

\section{$\CPN$ model}
The $\CPN$ model involves an $N$-component complex scalar field $\phi$, a $U(1)$ gauge field $A_\mu$, and a Lagrange multiplier field $\lambda$. The Lagrangian is \cite{wittenCPN, colemanBook, zinnJustin, largeN1, largeN3, pufu}
\begin{equation} \label{CPN}
	\L = \frac{N}{g} \Bigl\{ | \del \phi - i A\phi|^2 + \tfrac18 \mathcal R |\phi|^2 + \lambda\left( |\phi|^2 - 1 \right) \Bigr\},
\end{equation}
where $\mathcal R$ is the Ricci scalar. This theory describes a phase transition between a Higgs phase (at small $g$) and a Coulomb phase (at large $g$). We will study it at criticality $g=g_c$, at which point it is a CFT.

Given that $\phi$ only appears quadratically in \eqref{CPN}, it can be formally integrated out exactly. This leaves us with an effective action $S_\text{eff}[A,\lambda]$ for $A$ and $\lambda$:
\begin{equation}
	S_\text{eff}[A, \lambda] = N \tr\ln[-(\del - i A)^2 + \tfrac18 \mathcal R + \lambda ] -\frac N g \int  \lambda.
\end{equation}
Due to the overall factor of $N$ in front, this theory can be solved order by order in a large $N$ expansion. Expanding $A$ and $\lambda$ about a saddle point,
\begin{align}\label{background}
	A &= \bar A + a \\
	\lambda &= \mu^2 + \sigma,
\end{align}
gives
\begin{equation} \label{quadAction}
	S_\text{eff}[A, \lambda]  = S_\text{eff}[\bar A, \mu^2]   +  S_\text{eff}^{(2)}[a, \sigma]  + \cdots,
\end{equation}
where $S_\text{eff}^{(2)}[a, \sigma] $ is the quadratic effective action for the fluctuations $a$ and $\sigma$. Schematically, 
\begin{equation}
	S_\text{eff}^{(2)}[a, \sigma] =  \iint \Bigl[ \tfrac12 \vev{\mathcal J^\mu \mathcal J^\nu} a_\mu a_\nu +  \tfrac12 \vev{|\phi^2| |\phi^2|} \sigma \sigma  + \vev{\mathcal J^\mu |\phi|^2 } a_\mu \sigma \Bigr],
\end{equation}
where for compactness we suppressed spacetime arguments as well as a contact term (the ``seagull'' term familiar from scalar QED). The correlators are evaluated with $A$ and $\lambda$ treated as non-dynamical background fields fixed at $\bar A$ and $\mu^2$. The gauge current $\mathcal J_\mu$ is defined as
\begin{equation}
	\mathcal J_\mu = i \left[ \phi^\dag (\del_\mu-i \bar A_\mu) \phi - \phi (\del_\mu + i \bar A_\mu) \phi^\dag \right].
\end{equation}

\subsection{Saddle point and critical coupling}
We expect the lowest energy saddle point at fixed $Q$ to have a uniformly distributed magnetic field. Therefore, we take the background $\bar A$ to be
\begin{equation} \label{Abar}
	\bar A = \tfrac12Q(1-\cos \theta) d \phi,
\end{equation}
which results in the uniform field strength $d \bar A = \tfrac12Q\sin \theta d\theta\wedge d\phi$. 

The equation of motion for $\lambda$ imposes the constraint (after canonically normalizing $\phi$)
\begin{equation}
	\vev{|\phi|^2}_{Q,\mu^2}  = \frac{N}{g},
\end{equation}
where we explicitly indicated the dependence on the background values of $A$ and $\lambda$. For fixed~$N$ and $g$,  this is a constraint between $\mu^2$ and $Q$. In other words, for different values $\mu'^2$ and~$Q'$, we have 
\begin{equation} \label{constraint}
	\vev{|\phi|^2}_{Q,\mu^2} = \vev{|\phi|^2}_{Q',\mu'^2}.
\end{equation}
From \eqref{CPN} and \eqref{background}, we see that $\mu^2$ is the mass parameter for the $\phi$ field. To reach criticality, we tune $g$ so that $\mu^2$ vanishes when~$Q$ vanishes. This defines the leading order in large $N$ critical coupling $g=g_c$. From \eqref{constraint}, we have
\begin{equation} \label{constraint2}
	\vev{|\phi|^2}_{Q,\mu^2}  - \vev{|\phi|^2}_{0,0} = 0.
\end{equation}
By expanding the scalar field $\phi$ in terms of monopole spherical harmonics \cite{monopoleHarmonics}, 
\begin{equation} \label{monopoleHarmonics}
	\phi = \int \frac{d\omega}{2\pi} \sum_{j=Q/2}^\infty \sum_{m=-j}^j \Phi_{jm}Y_{Q/2,jm}(\theta, \phi)e^{-i\omega \tau},
\end{equation} 
\eqref{constraint2} can be evaluated and gives \cite{largeN1,largeN3,pufu}
\begin{equation} \label{mu}
	\sum_{j=Q/2}^\infty \left( \frac{j+\tfrac12}{\sqrt{(j+\tfrac12)^2+\mu^2-\tfrac14 Q^2}} -1 \right) - \tfrac12 Q = 0.
\end{equation}
For fixed $Q$, it is easy to numerically compute $\mu^2$ as the solution to this equation.  At leading order in large $N$, criticality is encoded in this specific relation between $\mu^2$ and $Q$.  At subleading order, $g_c$ receives a correction from the self energy of the $\phi$ field \cite{pufu, sachdev}.

\section{Computation}
The dimension $\Delta$ of the lowest dimension monopole operator is \eqref{stateOp}. To subleading order in~$1/N$, this is given by the path integral over the fluctuations $a$ and $\sigma$ weighted by the effective action~\eqref{quadAction} expanded to quadratic order. The result is
\begin{equation} \label{delta}
	\Delta =  \lim_{T\to \infty} \frac{1}{T} \left\{ S_\text{eff}[\bar A, \mu^2]  + 
	\tfrac12 \log \det\!' \begin{pmatrix}
		 \vev{|\phi^2| |\phi^2|} &  \vev{J^\mu |\phi|^2 } \\
		  \vev{J^\mu |\phi|^2 } &  \vev{J^\mu J^\nu} 
	\end{pmatrix} 
	\right\} + \O\left(\frac{1}{N} \right),
\end{equation}
where the prime on $\det \! '$ means to take the product of only the non-zero eigenvalues. Due to gauge invariance, there will be eigenvalues that vanish (see \cite{largeN6} for details on gauge fixing). The first term in \eqref{delta} is the leading $\O(N)$ contribution to~$\Delta$ and the second term is $\O(1)$ in~$N$. Let us write this as
\begin{equation}
	\Delta \equiv N \Delta_1 +  \Delta_0 + \cdots .
\end{equation}
Assuming the large charge expansion \eqref{largeQ}, we have
\begin{align} 
	\Delta_1 &= \alpha_1 Q^{3/2} + \beta_1 \sqrt Q + \gamma_1 + \cdots \label{N1}\\
	\Delta_0 &= \alpha_0 Q^{3/2} + \beta_0 \sqrt Q + \gamma_0 + \cdots. \label{N0}
\end{align}
In particular, $\gamma$ in \eqref{largeQ} has the expansion
\begin{equation}
	\gamma = N \gamma_1 + \gamma_0 + \frac{\gamma_{-1}}{N} + \cdots
\end{equation}
One immediate prediction of the large charge expansion is that all $\gamma_i$ vanish except for $\gamma_0$. In other words, $\gamma=\gamma_0$.
This follows from the prediction that $\gamma$ is universal and thus cannot depend on $N$.  In this paper, we verify that $\gamma_1 = 0$ and that $\gamma_0 = \gamma$ with $\gamma$ given by~\eqref{prediction}. We do not verify that $\gamma_i$ vanishes for $i<0$ since we only work to subleading order in large~$N$.

By expanding the scalar field $\phi$ in terms of monopole spherical harmonics as above \eqref{monopoleHarmonics},
the determinant \eqref{delta} was numerically evaluated for $Q = 1, 2, \dots,5$ in a computational tour de force \cite{pufu}. We simply repeated that computation for $Q = 1,2, \cdots, 100$. For all the details, see \cite{pufu}. Note that our definition of $Q$ differs from \cite{pufu} by a factor of 2.

\section{Results}
\subsection{Leading order}
\begin{figure}
	\centering
	\includegraphics[width=0.47\textwidth]{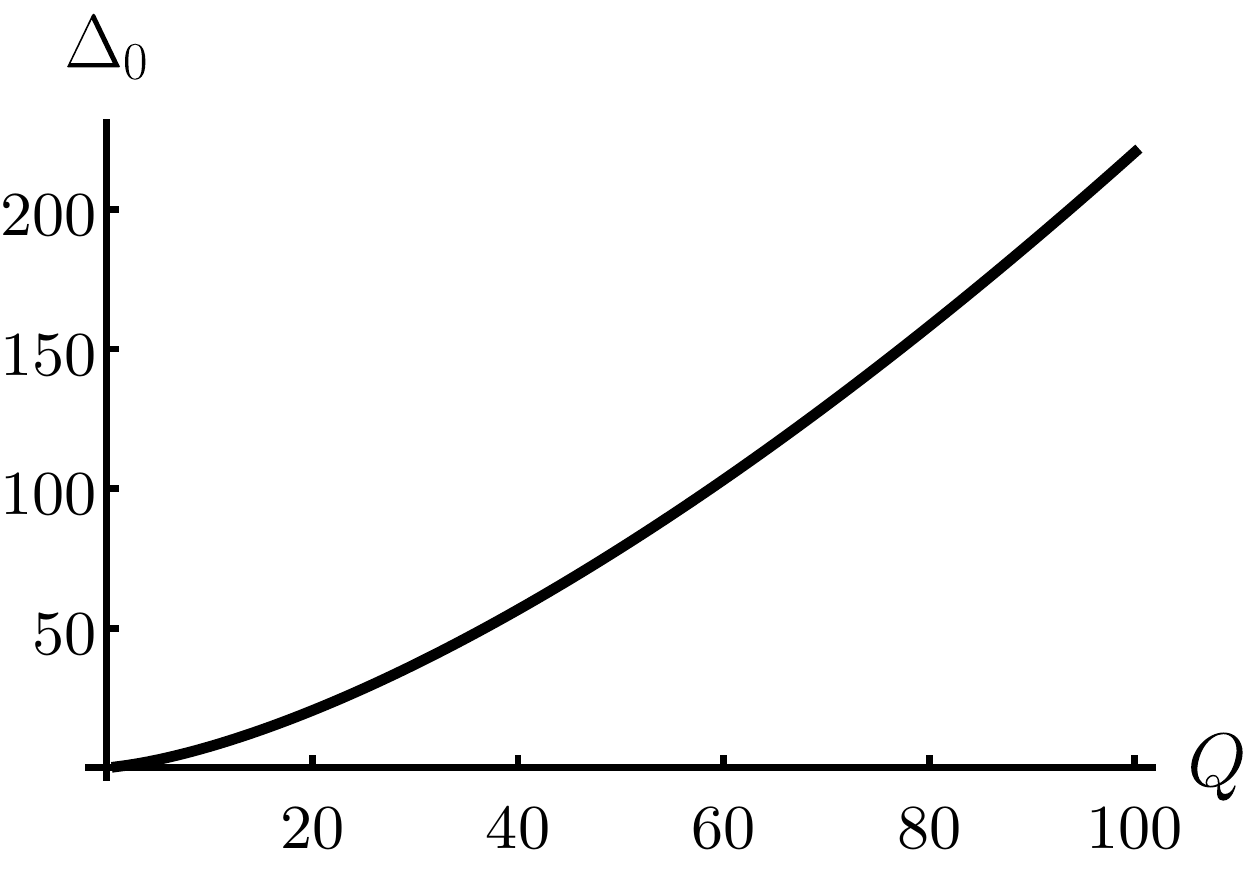}
	\captionsetup{width=.47\textwidth}
	\caption{The $\O(N^0)$ contribution to $\Delta$}
	\label{plt1}
\end{figure}

The saddle point contribution to the ground state energy $\Delta_1$ can be expressed as the convergent sum \cite{largeN3,pufu}
\begin{multline} \label{delta1}
	\Delta_1 = 2 \sum_{j=Q/2}^\infty \left[ (j+\tfrac12) \sqrt{(j+\tfrac12)^2 + \mu^2 -\tfrac14Q^2} - (j+\tfrac12)^2 - \tfrac{1}{2} (\mu^2 - \tfrac14Q^2) \right] \\
	-\frac{Q \mu^2}{2} + \frac{Q(1+ \tfrac12 Q^2)}{12},
\end{multline}
where $\mu^2$ is determined as the solution to \eqref{mu}. 

At large $Q$, the sum \eqref{delta1} for $\Delta_1$ can be analytically computed order by order in $1/Q$ \cite{pufu}, giving us the coefficients in \eqref{N1}:
\begin{align}
	\alpha_1 &= \zeta(-\tfrac12,\tfrac12 + \chi_0) \label{alpha1} = 0.09336639094\dots \\
	\beta_1 &= \tfrac52 \zeta(-\tfrac32,\tfrac12+\chi_0) -3 \alpha_1  \chi_0  = 0.03574650308\dots \\
	\gamma_1 &=0,
\end{align}
where $\chi_0 =-0.19727817140\dots$ is the solution to $\zeta(\tfrac12, \tfrac12 + \chi_0) = 0$ and $\zeta$ is the Hurwitz zeta function. As predicted,~$\gamma_1$ vanishes.  This really just follows from the fact that both $\mu^2/Q$ and $\Delta_1/Q^{3/2}$ have a power series expansion in $1/Q$.

\subsection{Subleading order}
We repeated the computation of \cite{pufu} for $Q = 1,2, \dots, 100$. The results are plotted in Figure~\ref{plt1} and the numerical values are displayed in Appendix~\ref{A:rawData}. By fitting to \eqref{N0}, we found that
\begin{equation} \label{fit}
	\Delta_0^\text{fit} = 0.2182275 \,Q^{3/2}+0.23764 \sqrt{Q} - 0.0935 + \frac{0.025}{\sqrt Q} + \cdots.
\end{equation}
In particular, our value for $\gamma$ is
\begin{equation} \label{finalAnswer}
	\gamma^\text{fit} =  - 0.0935 \pm 0.0003 ,
\end{equation}
which is both consistent with and within one percent of the value of $\gamma = -0.0937256\dots$ predicted in \cite{hellerman1, sasha}.
This is the main result of our paper. We will give details about our error analysis in the next three subsections.

\begin{figure}
        \centering
        \begin{subfigure}[t]{0.47\textwidth}
        		\vskip 0pt
                	\centering
                	\includegraphics[width=\textwidth]{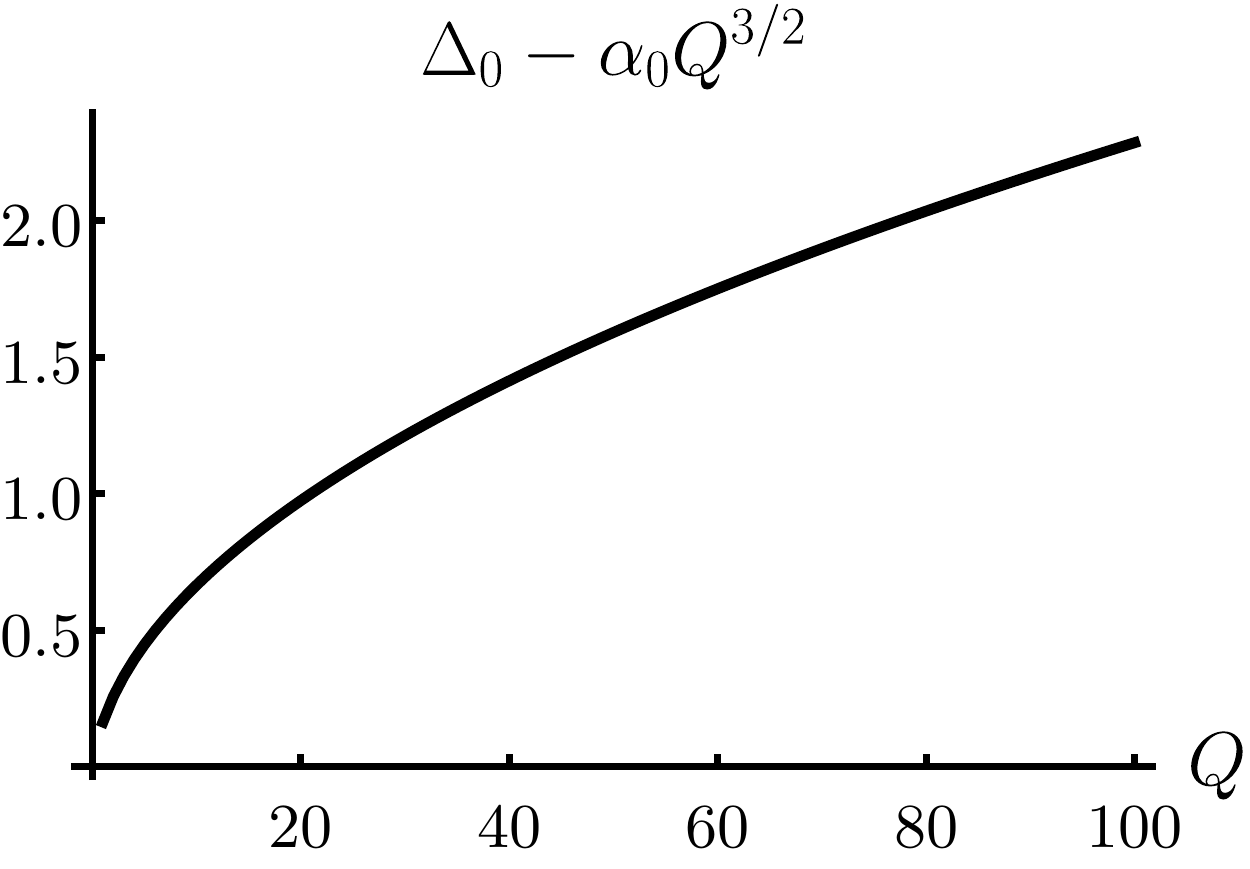}
                	\caption{Subtracting just the $\O(Q^{3/2})$ term of the fit~\eqref{fit} from the computed values of $\Delta_0$}
                	\label{plt2}
        \end{subfigure}
        \hfill
        \begin{subfigure}[t]{0.47\textwidth}
        		\vskip 0pt
                	\centering
                	\includegraphics[width=\textwidth]{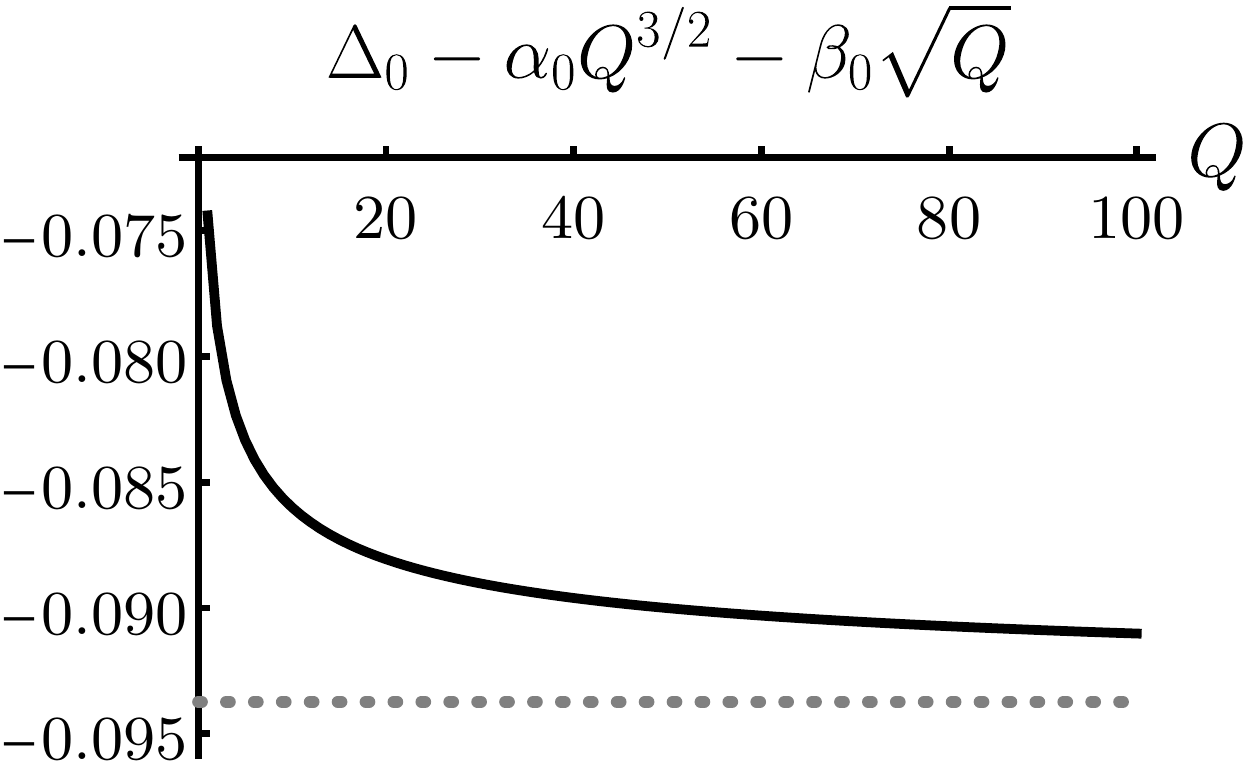}
		\vspace{.3mm}
                	\caption{Subtracting both the $\O(Q^{3/2})$ and $\O(\sqrt Q)$ terms. The dashed line is the predicted~$\O(Q^0)$ contribution \eqref{prediction}.}
                	\label{plt3}
        \end{subfigure}
        \caption{Comparing the fit \eqref{fit} with the computed values of $\Delta_0$ (shown in Figure~\ref{plt1})}
        \label{plt23}
\end{figure}

In our fit, we assumed the functional form of the expansion \eqref{largeQ}. To illustrate the validity of this form, Figure~\ref{plt23} plots the difference between our computed values of $\Delta_0$ and the fit~\eqref{fit}. In Figure~\ref{plt2}, the  subleading $\O(\sqrt Q)$ behavior is clearly visible in our computed values. In Figure~\ref{plt3}, the universal constant plus the remaining $\O(1/\sqrt Q)$ behavior is also visible.

\subsubsection{Numerical Approximations} \label{deltaError}
There are two places in the numerical computation where approximations are made. They both involve approximating a sum of an infinite number of terms. Let the summation variable be called~$n$. First, the sum is truncated at a large value $n_\text{max}$. Then, the summand is expanded in a~$1/n$ expansion up to $\O(1/n^p)$, and the remainder of the sum is analytically evaluated using this expansion. Finally,~$n_\text{max}$ and~$p$ are increased until the desired precision is reached. We computed each value of $\Delta$ to 6 decimal places. In practice, that meant increasing~$n_\text{max}$ and~$p$ until the numerical value of $\Delta$ stopped changing in the 6th decimal place. We then took our error bar for each $\Delta$ to be $10^{-6}$.

More specifically, we set $\Lambda = 70$ in (4.63) of \cite{pufu} and carried out the expansion (C.49) of \cite{pufu} to $\O\left( [(j+\frac12)^2+\omega^2]^{-6} \right)$. We also applied the same technique described in Figure 1 of \cite{largeN6} with $j_c =200$, and we carried out that expansion to $\O(j^{-18})$.

\subsubsection{Fitting Details} \label{fitError}
The value we obtain for the $\O(Q^0)$ piece depends on how many terms in the large $Q$ expansion we include in the fit. Let us denote the number of terms by $k$. In principle, $k = \infty$. However, if $k$ is too large, this will result in overfitting to the ``noise'' due to the finite precision of our numerical calculations. To determine an appropriate value for $k$, we borrowed a method that is common in machine learning. The 100 computed values of~$\Delta_0$ are randomly split into a~70-element ``training set'' and a 30-element ``cross validation'' set.  Then, the fit is performed on just the training set. Finally, $k$ is chosen to minimize the squared difference between the fit with $k$ terms and the cross-validation set. Repeating this hundreds of times, we found that $k = 9$ on average with a standard deviation of 2.

\subsubsection{Error analysis} \label{error}
The error of $\pm 0.0003$ in \eqref{finalAnswer} comes from combining the two sources of error mentioned in the last two subsections. There is an uncertainty in $k$ of $\pm 2$ and there is an uncertainty in~$\Delta$ of~$\pm 10^{-6}$. For each $k$ from 7 to 11, we propagated the uncertainty in $\Delta$ into an uncertainty in~$\gamma$, giving us an ``allowed region'' of $\gamma$ for each $k$. Then we took the total allowed region to be the union of the allowed regions.  Our reported error of $\pm 0.0003$ is the width of this region. 
\section{Discussion}
\subsection{Relation to the large charge EFT}
Physically, the background magnetic field gaps the $\phi$ and $\lambda$ fields by an amount of order~$\sqrt B$. Therefore, on distance scales larger than $1/\sqrt B$, we should be left with a local effective Lagrangian for the gauge field. This is just the Euler-Heisenberg effective Lagrangian on $S^2 \times \mathbb R$ with additional constraints due to conformal invariance. This EFT can be shown  \cite{hellerman1, vortex} to be equivalent to the large charge EFT. 

In this paper, we did our computations in the full UV-complete theory, not in the EFT. All divergences were absorbed by terms in the renormalizable Lagrangian \eqref{CPN} which is what allowed us to determine the non-universal coefficients $\alpha$ and $\beta$ in \eqref{largeQ}. 

\subsection{Flat space consistency check}
One simple consequence of the locality of the EFT is that the coefficient of the $\O(Q^{3/2})$ term in \eqref{largeQ} should be computable in flat space $\mathbb R^3$ without going to the cylinder $S^2 \times \mathbb R$.
Therefore, as a consistency check, we also repeated the entire computation of \cite{pufu} on $\mathbb R^3$. The main ingredient was the flat space $\phi$ propagator in a background magnetic field computed in \cite{schwinger}. Using this, we reproduced the numerical value of the $\O(N^0 Q^{3/2})$ coefficient in \eqref{fit} to 4 decimal places. (The numerical value of the $\O(N Q^{3/2})$ coefficient \eqref{alpha1} was already reproduced via a flat space calculation in \cite{pufu} and has a simple interpretation as a sum over Landau levels.)
%

\subsection{Outlook}
To summarize, using the methods of \cite{pufu}, we numerically computed the scaling dimensions of large charge monopole operators in the $\CPN$ model in 2+1 dimensions.  We computed the dimensions of the lowest dimension monopole operators of charge $Q = 1,2, \cdots 100$. From this, we fit to the large charge expansion of \cite{hellerman1} and verified the universal prediction for the $\O(Q^0)$ contribution.
In the future, it would be interesting to verify other predictions of the large charge EFT. In particular, there are specific predictions for the 4-point function \cite{riccardo, zhiboedov} as well as for the dimensions of operators with both large charge and large spin \cite{vortex}.
%

\section*{Acknowledgments}
We thank Gabriel Cuomo, M\'ark Mezei, Alexander Monin, and Riccardo Rattazzi for useful discussions and correspondence. The work of A.D.\ is partially supported by the Swiss National Science Foundation  under contract 200020-169696 and through the National Center of Competence in Research SwissMAP.


\appendix

%
%
%
%
%
%


\section{Numerical Values} \label{A:rawData}

\begin{center}
\begin{tabular}{cr}
\toprule
$Q$ & $\Delta_0$ \\
\midrule
 1 & 0.3814657 \\
 2 & 0.8745160 \\
 3 & 1.4645878 \\
 4 & 2.1387552 \\
 5 & 2.8878980 \\
 6 & 3.7052725 \\
 7 & 4.5856569 \\
 8 & 5.5248609 \\
 9 & 6.5194280 \\
 10 & 7.5664451 \\
 11 & 8.6634141 \\
 12 & 9.8081621 \\
 13 & 10.9987765 \\
 14 & 12.2335573 \\
 15 & 13.5109803 \\
 16 & 14.8296687 \\
 17 & 16.1883708 \\
 18 & 17.5859425 \\
 19 & 19.0213322 \\
 20 & 20.4935694 \\
 21 & 22.0017546 \\
 22 & 23.5450511 \\
 23 & 25.1226780 \\
 24 & 26.7339043 \\
 25 & 28.3780437 \\
 \bottomrule
\end{tabular}
\qquad
\begin{tabular}{cr}
\toprule
$Q$ & $\Delta_0$ \\
\midrule
 26 & 30.0544503 \\
 27 & 31.7625150 \\
 28 & 33.5016616 \\
 29 & 35.2713444 \\
 30 & 37.0710456 \\
 31 & 38.9002727 \\
 32 & 40.7585567 \\
 33 & 42.6454502 \\
 34 & 44.5605261 \\
 35 & 46.5033756 \\
 36 & 48.4736074 \\
 37 & 50.4708459 \\
 38 & 52.4947311 \\
 39 & 54.5449166 \\
 40 & 56.6210693 \\
 41 & 58.7228684 \\
 42 & 60.8500049 \\
 43 & 63.0021804 \\
 44 & 65.1791070 \\
 45 & 67.3805065 \\
 46 & 69.6061098 \\
 47 & 71.8556564 \\
 48 & 74.1288944 \\
 49 & 76.4255792 \\
 50 & 78.7454740 \\
 \bottomrule
\end{tabular}
\qquad
\begin{tabular}{cr}
\toprule
$Q$ & $\Delta_0$ \\
\midrule
 51 & 81.0883489 \\
 52 & 83.4539806 \\
 53 & 85.8421523 \\
 54 & 88.2526533 \\
 55 & 90.6852786 \\
 56 & 93.1398288 \\
 57 & 95.6161099 \\
 58 & 98.1139328 \\
 59 & 100.6331133 \\
 60 & 103.1734718 \\
 61 & 105.7348333 \\
 62 & 108.3170269 \\
 63 & 110.9198859 \\
 64 & 113.5432476 \\
 65 & 116.1869529 \\
 66 & 118.8508466 \\
 67 & 121.5347767 \\
 68 & 124.2385947 \\
 69 & 126.9621556 \\
 70 & 129.7053171 \\
 71 & 132.4679403 \\
 72 & 135.2498889 \\
 73 & 138.0510297 \\
 74 & 140.8712321 \\
 75 & 143.7103681 \\
 \bottomrule
\end{tabular}
\qquad
\begin{tabular}{cr}
\toprule
$Q$ & $\Delta_0$ \\
\midrule
 76 & 146.5683122 \\
 77 & 149.4449416 \\
 78 & 152.3401356 \\
 79 & 155.2537759 \\
 80 & 158.1857465 \\
 81 & 161.1359335 \\
 82 & 164.1042251 \\
 83 & 167.0905115 \\
 84 & 170.0946850 \\
 85 & 173.1166398 \\
 86 & 176.1562717 \\
 87 & 179.2134786 \\
 88 & 182.2881600 \\
 89 & 185.3802172 \\
 90 & 188.4895531 \\
 91 & 191.6160722 \\
 92 & 194.7596805 \\
 93 & 197.9202857 \\
 94 & 201.0977969 \\
 95 & 204.2921245 \\
 96 & 207.5031806 \\
 97 & 210.7308784 \\
 98 & 213.9751325 \\
 99 & 217.2358589 \\
 100 & 220.5129749 \\
 \bottomrule
\end{tabular}
\end{center}

\bibliographystyle{utphys}
\bibliography{references} 
\end{document}